\documentclass[a4paper]{jpconf}
\usepackage[utf8]{inputenc}
\usepackage{amsmath}
\usepackage{amssymb}
\usepackage{color}
\usepackage{graphicx}
\usepackage{multirow}
\usepackage{todonotes}
\usepackage{braket}
\begin{document}
\title{Investigation of $B\bar B$ four-quark systems using lattice QCD}

\author{Antje Peters$^1$, Pedro Bicudo$^2$, 
Krzysztof Cichy$^{1,3}$, Marc Wagner$^1$}

\address{$^1$ Goethe-Universit\"at Frankfurt am Main, Institut f\"ur Theoretische Physik,
Max-von-Laue-Stra{\ss}e 1, D-60438 Frankfurt am Main, Germany}
\address{$^2$ CFTP, Departamento de Física, Instituto Superior Técnico, Universidade de Lisboa,
Avenida Rovisco Pais, 1049-001 Lisboa, Portugal}
\address{$^3$ Adam Mickiewicz University, Faculty of Physics,
 Umultowska 85, 61-614 Poznan, Poland}

\ead{peters@th.physik.uni-frankfurt.de}

\begin{abstract}
We investigate $B \bar B$ systems by computing potentials of two static quarks in the presence of two quarks of finite mass using lattice QCD. By solving the Schrödinger equation we check whether these potentials are sufficiently attractive to host bound states. Particular focus is put on the experimentally most promising bottomonium-like tetraquark candidate $Z_b^\pm$ with quantum numbers $I(J^P)=1(1^+)$.
\end{abstract}

\section{Motivation}
Quite a number of mesons observed experimentally (e.g.\ at LHCb and Belle) are theoretically not well understood yet. One example is the charged bottomonium-like $Z_b^\pm$ (cf.\ e.g.\ \cite{Belle:2011aa}). It contains a $b\bar b$ pair, which one can infer from both its mass and decay products. Because it carries electromagnetic charge, there must be additional quarks, most likely a light quark-antiquark pair, i.e.\ the minimal quark content of $Z_b^\pm$ is $b\bar b u\bar d$/$b \bar b d \bar u$. Such four-quark states are of particular interest because most observed mesons can be described within the quark model using only a quark-antiquark pair. For this reason, four-quark states are the subject of many recent experimental and theoretical investigations.

In the following, we investigate the $Z_b^\pm$ assuming a four-quark structure $b\bar b u\bar d$ using lattice QCD. We approximate the heavy quarks $b$ and $\bar b$ by infinitely heavy static quarks $Q$ and $\bar Q$. We compute the potentials of the static quarks in the presence of the light quarks, which can be interpreted as the potentials of a $B$ and a $\bar B$ meson. To check whether the four quarks may form a bound state, we use the Born-Oppenheimer approximation, i.e.\ we solve the Schrödinger equation with the computed potentials.

For similar studies of static-light $B B$ systems within the same lattice QCD setup, cf.\ \cite{Wagner:2010ad,Wagner:2011ev,Bicudo:2012qt, Wagenbach:2014oxa, Bicudo:2015vta, Scheunert:2015pqa, Peters:2015tra, Bicudo:2015kna}. Other papers studying pairs of static-light mesons with similar methods include \cite{Stewart:1998hk,Michael:1999nq,Cook:2002am,Doi:2006kx,Detmold:2007wk,
Bali:2010xa,Brown:2012tm}.

\section{The $b\bar b u\bar d$ four-quark system -- qualitative discussion and expectations}

The $b\bar b u\bar d$ four-quark system can be characterized by the separation $r$ of the static quark $b \equiv Q$ and the antiquark $\bar b \equiv \bar Q$, by parity $P$, total angular momentum $J$ and isospin $I$. Since we consider static $b$ quarks, i.e.\ $b \equiv Q$ and $\bar b \equiv \bar Q$, the $b \bar b$ separation is also a quantum number. Moreover, since the static quark spin decouples from the Hamiltonian, both the light total angular momentum $j$ and the static quark spin $j_b$ ($j + j_b = J$) are separately conserved.

An experimentally very interesting case is the recently measured $Z_b^\pm$ with quantum numbers $I(J^P)=1(1^+)$. Isospin $I=1$ and positive electric charge is realised by light quark flavours $u\bar d$. Parity $P=+$ is consistent with a possible loosely bound $B \bar B$ structure, since both $B$ and $\bar B$ have $P = -$ and hence in combination result in $P = +$. Numerically, we find most evidence for a four-quark bound state when total light angular momentum $j = 0$ (cf.\ section~\ref{sec:comp}). Due to degeneracy with respect to the static quark spin $j_b$, the total angular momentum can either be $J = 0$ or $J = 1$, i.e.\ all our statements apply to a $b\bar b u\bar d$ four-quark system not only with $I(J^P)=1(1^+)$, but also with $I(J^P)=1(0^+)$.

States within the $I(J^P)=1(1^+)$ sector might have different structures including the following.
\begin{itemize}
\item A bound four-quark state made of a loosely bound $B \bar B$ meson pair (a so-called mesonic molecule), referred to as $B \bar B$.

\item A bound four-quark state made of a diquark and an anti-diquark.

\item A two-meson state made of a $B$ meson and a far separated $\bar B$ meson. Because the structure of the corresponding four-quark creation operator is the same as for the loosely bound $B \bar B$ meson pair (cf.\ eq.\ (\ref{EQN001}) below), this case is also referred to as $B \bar B$.

\item A two-meson state made of a bottomonium state and a far separated pion $\pi^+$. In the static approximation, a bottomonium state is realized by the static quark $Q$ and the static antiquark $\bar{Q}$ connected by a gluonic string. We refer to this case as $Q \bar Q+\pi$.
\end{itemize}

At small separations $r$, the ground state is $Q\bar Q +\pi$, while at large separations $r$ the ground state is $B \bar B$. Using lattice QCD we investigate in the next section whether for small separations $r$ there is a first excitation below the $B \bar B$ threshold that has a structure dominated by a four-quark bound state. A first excitation with such properties is an important prerequisite for a $b\bar b u\bar d$ bound state. Whether such a $b\bar b u\bar d$ bound state does indeed exist requires further analysis also discussed in the next section (using the Born-Oppenheimer approximation, solving the Schr\"odinger equation).

\section{\label{sec:comp}Lattice QCD computation of the $Q \bar Q$ potential in the presence $u\bar d$}

We start by discussing suitable creation operators $\mathcal O_j$ that generate field excitations, which are similar to the four-quark states of interest. The correlation functions of these operators at large temporal separation provide low-lying masses, which can be interpreted as potentials $V_0(r)$, $V_1(r)$, ..., because the operators $\mathcal O_j$ and thus also the masses depend on the $Q\bar Q$ separation $r$. The correlation functions read
\begin{equation}
\label{corr} C_{j k}(t)=\bra \Omega \mathcal O_j^\dagger(t) \mathcal O_k(0) \ket \Omega \underset{\textrm{large } t}{\approx}A^0_{j k} \exp{(-V_0(r)t)} + A^1_{j k}\exp{(-V_1(r)t)} + \ldots.
\end{equation}
The creation operators we have currently implemented are one that excites a $B \bar B$ state and another that excites a $Q \bar Q + \pi$ state:
\begin{align}
 &\mathcal O_0\equiv \mathcal O_{B\bar B}=\Gamma_{AB} \tilde \Gamma_{CD}\bar Q^a_C(\vec x) q^{a}_A(\vec x) \bar q^{b}_B(\vec y) Q^b_D(\vec y), \label{EQN001}\\
 &\mathcal O_1\equiv \mathcal O_{Q\bar Q+\pi}= \bar Q^a_A(\vec x) U^{ab}(\vec x, t; \vec y, t)\tilde \Gamma_{AB} Q^b_B(\vec y)\sum\limits_{\vec z} \bar q^{c}_C(\vec z)\left(\gamma_5\right)_{CD} q^{c}_D(\vec z), \label{QQpi}
\end{align}
with $|\vec x - \vec y|=r$. $\tilde\Gamma$ appearing in both operators is a combination of Dirac matrices that realizes either $j_b=0$ or $j_b=1$. It does not affect the potentials $V_j(r)$ since the spin of the heavy quarks is irrelevant as explained above. The matrix $\Gamma$ is a combination of Dirac matrices leading to the same quantum numbers $(j_z , P\circ C , P_x)$ as the $Q \bar Q + \pi$ operator \eqref{QQpi} (for details on ($j_z$, $ P$, $ C$, $P_x$), cf. e.g.\ \cite{Bicudo:2015kna}). Among the possible choices for $\Gamma$, the combination $\Gamma = \gamma_5 - \gamma_0 \gamma_5$ yields the strongest $Q \bar Q$ attraction, if one takes into account only the operator $\mathcal O_{B\bar B}$. Therefore, we consider this combination as the most promising for further analysis. $U^{ab}(\vec x, t; \vec y, t)$ denotes a product of gauge links connecting the two static quarks.

At the moment, we have performed computations of the correlation functions $C_{j k}(t)$ only on a single enesemble of gauge link configurations generated by the European Twisted Mass Collaboration (ETMC) with 2 dynamical quark flavours (for details cf.\ Table \ref{tab:ETMC}). To improve statistical accuracy, we have averaged these correlation functions with respect to the symmetries: time-reversal, parity, charge conjugation, $\gamma_5$-hermiticity and cubic rotations.

\begin{table}\begin{center}
\begin{tabular}{c|c|c|c|c|c}
$\beta$ & lattice size & $a \mu_l$ & $a$ in $\textrm{fm}$ & $m_\pi$ in $\textrm{MeV}$ & \# of configurations\\ 
\hline 
3.90 & $24^3\times 48$ & 0.0085 & 0.079 & 480 & 100 \\ 
\end{tabular} 
\caption{\label{tab:ETMC}Parameters of the gauge link configurations; $\beta$: inverse gauge coupling; $a \mu_l$: bare $u/d$ quark mass; $a$: lattice spacing; $m_\pi$: pion mass.}
\end{center}\end{table}

In particular, the extraction of the energy of the first excited state $V_1(r)$ is a numerically rather challenging task, which we do by solving a generalized eigenvalue problem (GEP) (cf.\ \cite{Blossier:2009kd} and references therein).

\begin{center}
\begin{figure}
\includegraphics[width=\textwidth]{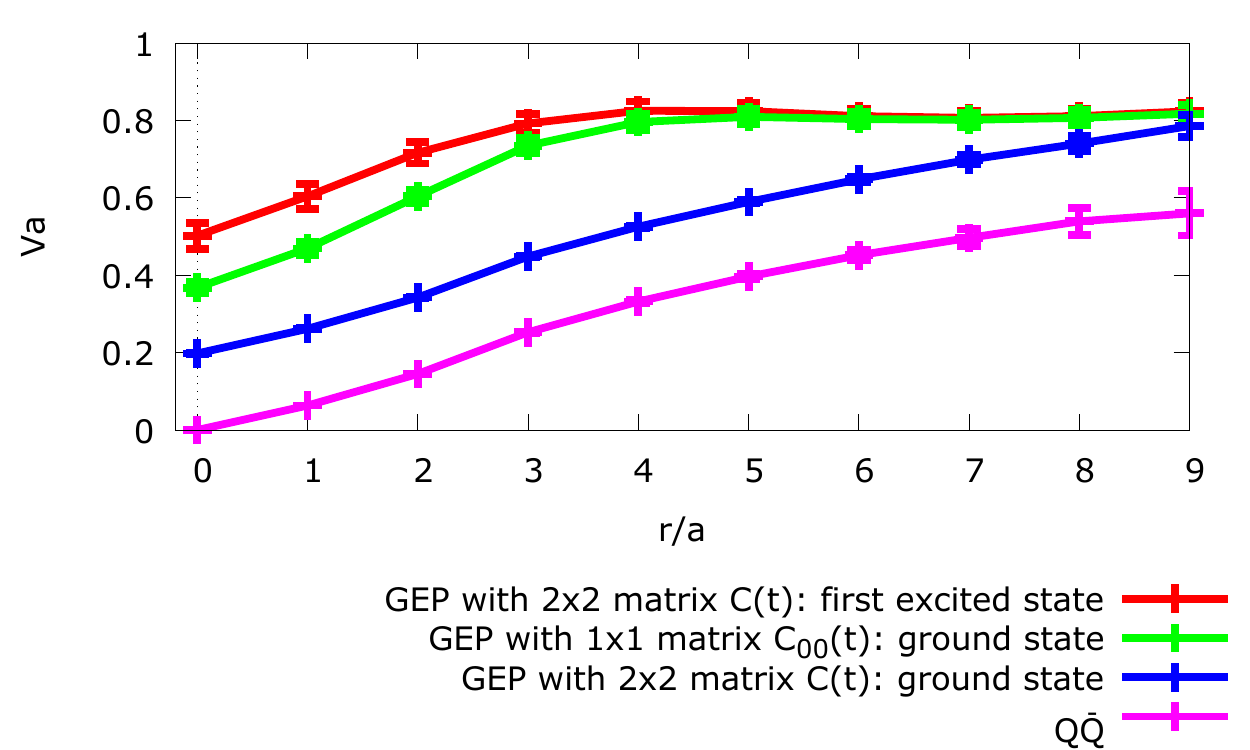}
\caption{\label{pic:2x2}$Q\bar Q u\bar d$ potentials and the $Q\bar Q$ potential (cf.\ the text for details).}
\end{figure}
\end{center}

Figure~\ref{pic:2x2} shows $Q\bar Q u\bar d$ potentials obtained via the GEP from our correlation matrix (\ref{corr}) as well as the ordinary static $Q\bar Q$ potential (obtained from standard Wilson loops). The blue curve corresponds to the ground state $V_0(r)$ (obtained from the full $2 \times 2$ matrix). By comparing with the ordinary static $Q\bar Q$ potential, one can clearly see that it is just shifted by the pion mass $a m_\pi \approx 0.2$) indicating that the ground state of our four quark system is of the $Q \bar Q+\pi$ type (this is also supported by the eigenvectors obtained from the GEP). While the red curve corresponds to the first excitation $V_1(r)$ (obtained from the full $2 \times 2$ matrix), the green curve is the result obtained from a single correlation function, the correlation function of the operator $O_{B\bar B}$, i.e.\ from $C_{0 0}(t)$. The fact that the green curve is similar to, but somewhat below the red curve suggests that the first excitation $V_1(r)$ is predominantly of $B\bar B$ type (the operator $O_{B\bar B}$ seems to generate mostly the first excitation, but also a non-vanishing ground-state overlap). This attractive potential $V_1(r)$ is a potential candidate to host a four-quark bound state.

\section{Solving the Schrödinger equation to check for a bound state}

To check for a four-quark bound state and, if present, to obtain its binding energy $-E_B$, we solve the $s$-wave Schrödinger equation
\begin{equation}
\label{schroedinger} \bigg(-\frac{1}{2 \mu} \frac{d^2}{dr^2} + U(r)\bigg) R(r) = E_B R(r), \quad \textrm{with} \quad U(r) = V(r)\Big|_{V_0 = 0} \quad \textrm{and} \quad \mu = m_b/2 .
\end{equation}
$V$ denotes an analytic parameterization of our lattice QCD result for $V_1(r)$ (red points in Figure~\ref{pic:2x2}) obtained by fitting
\begin{equation}
\label{ansatz} V(r) = V_0 - \frac{\alpha}{r} \exp(-(r/d)^2)
\end{equation}
with respect to $V_0$, $\alpha$ and $d$. The existence of a four-quark bound state, i.e.\ a tetraquark, is indicated by $E_B < 0$. For more details on the extraction of the binding energy in this so-called Born-Oppenheimer approximation, cf.\ \cite{Peters:2015tra}.

Determining $V_1(r)$ from the GEP using different $t$ ranges and performing several fits of (\ref{ansatz}) with different $r$ ranges allows to estimate both $E_B$ and and its uncertainty,
\begin{equation*}
E_B =(-58 \pm 71) \, \textrm{MeV} .
\end{equation*}
This negative value provides first indication of the existence of an $I(J^P)=1(1^+)$ $b\bar b u\bar d$ tetraquark using lattice QCD with infinitely heavy $b$ quarks and the Born-Oppenheimer approximation. Our findings are consistent with the experimentally observed $Z_b^\pm$ states.

\section{Outlook}

The analysis presented above is very preliminary. In the near future, we plan for example to increase statistical precision and to extend our computations to lighter $u/d$ quark masses. Moreover, it will be interesting to explore also other $I(J^P)$ sectors. Finally, the spin of the heavy quarks should be included as e.g.\ pioneered in \cite{Scheunert:2015pqa}.

\section*{Acknowledgments}

P.B.\ thanks IFT for hospitality and CFTP, grant FCT UID/FIS/00777/2013, for support. A.P.\ and M.W.\ acknowledge support by the Emmy Noether Programme of the DFG (German Research Foundation), grant WA 3000/1-1.

This work was supported in part by the Helmholtz International Center for FAIR within the framework of the LOEWE program launched by the State of Hesse.

Calculations on the LOEWE-CSC high-performance computer of Johann Wolfgang Goethe-University Frankfurt am Main were conducted for this research. We would like to thank HPC-Hessen, funded by the State Ministry of Higher Education, Research and the Arts, for programming advice.

\end{document}